\documentclass[pra,aps,final,twocolumn,nofootinbib]{revtex4-1}
\usepackage{epsfig}
\usepackage{latexsym}
\usepackage{xspace}
\usepackage{hyperref}
\usepackage[latin2]{inputenc}
\usepackage{indentfirst}
\usepackage{enumerate}
\usepackage{color}
\usepackage{colordvi}

\usepackage{amsmath}
\usepackage{amssymb}
\usepackage[english]{babel}
\usepackage{url}

\def\be{\begin{equation}}
\def\ee{\end{equation}}
\def\bea{\begin{eqnarray}}
\def\eea{\end{eqnarray}}
\def\l{\label}

\def\p{{\bf p}}

\def\r{{\bf r}}

\def\u{{\bf u}}

\def\d{\mbox{d}}

\def\siml{\;\hbox{\kern.1em \lower.7ex \hbox{$\sim$} \kern-1.12em
 \raise.5ex \hbox{$<$} \kern.1em}}
\def\simg{\;\hbox{\kern.1em \lower.7ex \hbox{$\sim$} \kern-1.12em
 \raise.5ex \hbox{$>$} \kern.1em}}

\begin{document}

\title{Ultrasonic waves in classical gases}

\author{ A.G.\ Magner}
\affiliation{Institute for Nuclear Research NASU, 03680 Kiev, Ukraine}

\author{M.I.\ Gorenstein}
\affiliation{
Bogolyubov Institute for Theoretical Physics, 03143 Kiev, Ukraine}

\author{U.V.\ Grygoriev}
\affiliation{
Department of Physics, Taras Shevchenko National University of Kiev, 03022 Kiev, Ukraine}

\begin{abstract}
The velocity and absorption coefficient
for the plane sound waves in a classical gas
are obtained by solving the Boltzmann kinetic equation,
which describes the reaction of the single-particle distribution function
to  a periodic external field.
Within the linear response theory, the nonperturbative
dispersion equation valid for all sound frequencies is derived
and solved numerically. The
results are in agreement with the approximate
analytical solutions
found for both the frequent- and rare-collision regimes.
\end{abstract}

\maketitle
\centerline\today

\section{Introduction}

Sound waves in classical gases
have been studied intensively within the hydrodynamical approach
(see, e.g., Ref.~\cite{LLv6}).
Small dynamical perturbations
of the particle density $n(\r,t)$, collective velocity
$\u(\r,t)$, and temperature $T(\r,t)$
induced by sound-wave propagation are found
as solutions of the hydrodynamical and transport equations.
The sound velocity is approximately equal to the thermal
particle velocity\footnote{We use
the units where the Boltzmann constant is $k_B=1$.}
\be\l{vT}
v_T~=~\sqrt{\frac{2\,T}{m}}~,
\ee
where $T$ is the system temperature
and $m$ is the particle mass.
For absorbed plane sound waves (APSWs) with
a frequency $\omega$, the wave amplitude  decreases  as
$\exp(-\gamma z)$ after propagating the distance $z$.
The absorption coefficient
$\gamma$
is obtained from the Stokes relation \cite{LLv6}.
It is a function of the shear
and bulk
viscosity and the thermal conductivity.

Within a hydrodynamic approach, the kinetic coefficients
are phenomenological constants.
For their calculations,
one needs kinetic theory.
For classical systems of particles,
the Boltzmann kinetic equation (BKE) is usually used.
It describes  the single-particle  distribution function
$f(\r,\p,t)$ dependent on  the space coordinate $\r$,
momentum $\p$, and time $t$
(see, e.g.,
Refs.~\cite{chapman,kogan,silin,LPv10,reichl,cercignani,gorenstein}).
The two lowest
moments of the single-particle distribution give
\bea\l{p-mom}
n(\r,t)~&=&~\int d\p~f(\r,\p,t)~,\nonumber\\
\u(\r,t)~&=&~\frac{1}{n(\r,t)}\,\int d\p~\frac {\p}{m}~f(\r,\p,t)~.
\eea

Chapman and Enskog (see, e.g., Ref.~\cite{chapman})
derived the equations of a dissipative hydrodynamics
by using the BKE.
They obtained the expressions for the kinetic coefficients
 in the system of hard balls
within the so-called frequent-collision regime (FCR)
(see also Ref.~\cite{MGGPprc16}). To define
the collision regimes, let us
introduce two different scales: the particle mean free
path $l\sim (n\pi d^2)^{-1}$, where $d$ is the hard-core particle
diameter, and another scale $L$ that is the characteristic space
size of the external
dynamical perturbation. In our APSW problem, one can set $L=\lambda$,
where $\lambda =2\pi\,v^{}_T/\omega$
is the wavelength of the propagating plane waves.
The FCR corresponds to $l\ll \lambda$  and
the perturbation expansion over a small Knudsen parameter
${\mathcal K}\equiv \omega\tau \ll 1$
can be used. Here, $\tau \sim l/v_T$
is the relaxation time
that determines the collision frequency $\tau^{-1}$
by the
collision integral term in the BKE. In most practical cases, the
inequality $l \ll \lambda$ is satisfied,
and the FCR works (see, e.g., Ref.~\cite{LPv10}).
For example, for
air at normal conditions one has
$l\sim 10^{-5}$cm and $\lambda\sim  1 - 10^3$cm
for audible sound waves.
(For investigations within the FCR but
beyond the standard hydrodynamic approach; see,
for instance,
Refs.~\cite{Gr65book,kogan,cercignani,abrkhal,Gr65book,Si65,GrJa59,BrFe66,brooksyk,Le78,Wo79,Wo80,relkinbook,woodsbook93,Le89,Ha01,spiegel_Ultrasonic_01,spiegel_BVKE-visc_03,smith05,AWprc12,AWcejp12,wiranata-jp-14,kapusta}.)

The rare-collision regime (RCR)
takes place
at  large values of the Knudsen
parameter
${\mathcal K}=\omega\tau \gg 1$.
The conditions of the RCR emerge at a small
particle-number density ($l$ becomes large)
and  (very) large sound-wave frequency $\omega$.
Different approximations
were employed in this case
\cite{kogan,silin,cercignani,Si65,GrJa59,BrFe66,brooksyk,Le78,Wo79,Wo80,relkinbook,bhatia,Le89,woodsbook93,magkohofsh,Ha01,spiegel_Ultrasonic_01,pethsmith02,spiegel_BVKE-visc_03,smith05,KMprc06,litovitz,review,BMRprc15}.
Most of them, e.g., \cite{kogan,cercignani,Si65,BrFe66,brooksyk,Le78,Wo79,Wo80,relkinbook,Le89,woodsbook93,Ha01,spiegel_Ultrasonic_01,spiegel_BVKE-visc_03}, used
the so-called moments' method  based
on a truncation of the system of equations for the
moments of the BKE. However,
a dynamical variation $\delta f(\r,\p,t)$
of the distribution function $f(\r,\p,t)$ becomes strongly oscillating
and a nonsmooth
function of the momentum $\p$ at large $\omega\tau \simg 1$. Therefore,
in the RCR at $\omega\tau \simg 1$ the moments'
method fails,
the worse the larger $\omega\tau$ (see, e.g.,
Refs.~\cite{cercignani,relkinbook}).

The APSWs in the RCR will be referred to
as ultrasonic waves \cite{litovitz,bhatia}.
The basic experiments in the field of ultrasonic waves were done by
Greenspan (see Refs.~\cite{Gr49,Gr56,Gr65book},
and also Ref.~\cite{Me57} for additional experimental data).
After that no essential improvements of experimental
measurements have been done.
This is connected with serious difficulties in conducting
 these experiments. The difficulties include
the problems of generating ultrasonic waves in gases and of measuring the speed and absorption
of these waves. 

Both the FCR and RCR have been studied in our recent paper \cite{MGGpre17}
by using the approximate analytical solutions obtained
within asymptotic expansions
of the BKE over $\omega\tau \ll 1$ and $(\omega\tau)^{-1} \ll 1$, respectively.
The aim of the present paper is to formulate the nonperturbative
method for calculations of the speed of sound waves and  the
absorption coefficient
$\gamma$ of the APSWs
within the linear response theory (LRT) (for different aspects
of the LRT  see, e.g., Refs.~\cite{kubo,zubarev,hofmann}, and also
Refs.~\cite{magkohofsh,review} for its applications).
The LRT allows us to derive a general APSW solution
for small variations $\delta f(\r,\p,t)$
of the distribution function $f$ in terms of the response of $\delta f$
to a periodic external field with a frequency $\omega$
by using the relaxation time approximation
to the collision integral term.
This can be done
for any Knudsen parameter value $\mathcal{K}$,
which is an
essential advantage over the
moments' method.
The equation for the
poles of these response
functions is the dispersion equation for the complex wave number
$k$ (or the complex sound velocity),
which allows us to obtain the
sound velocity and the absorption coefficient.

The paper is organized as follows.
In Sec.\ II, the Boltzmann kinetic
equation with  an external periodic field is formulated,
and the relaxation time approximation for
the collision integral
is discussed.
In Sec.\ III, the linearized BKE is solved in terms of the
response  to the external periodic potential.
The  nonperturbative
dispersion equation for the complex wave numbers
(sound velocities) of the APSWs
is derived.
Numerical solutions of this equation give the sound
velocity and absorption coefficient of the APSWs
at arbitrary values of the
Knudsen parameter $\omega\tau$.
The obtained numerical results are in agreement with
the analytical RCR and FCR asymptotic limits.
Sections \ IV and V
present, respectively, a discussion of the results
and a summary. Appendixes \ref{appA}-\ref{appC}
show some details of
the calculations.

\section{Boltzmann Kinetic equation }
\l{sec-kintheor}

We start with the BKE
\be\l{BKE}
\frac{\partial f}{\partial t}~+~
\frac{{\bf p}}{m} \frac{\partial f}{\partial {\bf r}}
~- \mbox{St}[f]~=~
\frac{\partial f}{\partial \p}\;
\frac{\partial V_{\rm ext}}{\partial \r}\;,
\ee
where the collision  integral term $\mbox{St}[f]$
is taken in the standard Boltzmann
form (see, e.g., Refs.~\cite{kogan,silin,cercignani,reichl}).
The external potential field
$V_{\rm ext}(z,t)$, periodic in time $t$ with a frequency
$\omega$, is switched on as a
perturbation
\be\label{Vext}
V_{\rm ext}(z,t)=
\exp\left[-i \, \omega \,
t +\epsilon^{}_0t\right]
\int_{-\infty}^\infty \frac{\d k}{2 \pi}~ V_k \;
\exp\left(i k z\right)\;,
\ee
where $V_k$ is the $k$ amplitude of the Fourier
representation $\epsilon^{}_0=+0$.
The external field\footnote{As usual, the complex number representation is used for convenience,
but only the real parts of $f$ and $V_{\rm ext}$ will be taken as physical quantities.}
stimulates the appearance of the plane waves with
a fixed frequency $\omega$. The term $\epsilon^{}_0t$, switching
adiabatically on the
external field at a time far in the past
($V_{\rm ext}\rightarrow 0$ at $t \rightarrow -\infty$) , is used usually
for the adequate time-dependent picture \cite{kubo,zubarev}, and will
be omitted in the following derivations.
For convenience, we use also the standard Fourier integral
transformation of the LRT from the
$z$ to $k$ variables.

In the absence of the external field  $V_{\rm ext}$,
the global
equilibrium (GE) solution of the BKE (\ref{BKE}) is given by
\be\l{maxwell}
f^{}_{\rm GE}(p)~=~
\frac{n}{(2\pi mT)^{3/2}}\,
\exp\left(-\;\frac{p^2}{2mT}\right)\;,
\ee
where the particle number density $n$ and temperature $T$
are independent of the spatial coordinates
$\r$ and time $t$, $p\equiv |\p|$. Therefore,
Eq.~(\ref{maxwell}) for $f_{\rm GE}$ describes
the homogeneous particle distribution in the coordinate space and the
Maxwell distribution in the momentum space.
This distribution satisfies
Eq.~(\ref{p-mom}) with $n(\r,t) =n$ and $\u(\r,t)=0$.

Let us define now small deviations $\delta f(\r,\p,t)$ from the
distribution function $f_{\rm GE}(p)$
($|\delta f|/f_{\rm GE}\ll 1$),
\be\l{dfdef}
\delta f(\r,\p,t)~\equiv~f(\r,\p,t)~-~f^{}_{\rm \rm GE}(p)\;.
\ee
They arise from a small
external potential $|V_{\rm ext}|/T \ll 1~$.
The linearized BKE (\ref{BKE})
can be then written as
\be\l{}
\frac{\partial \delta f}{\partial t}+
\frac{{\bf p}}{m} \frac{\partial \delta f}{\partial {\bf r}}
-\delta \mbox{St}[f]~~~~~~~~~~~~~~~~~~~~~~~~~~~~~~~~~~~ \nonumber
\ee

\vspace{-0.7cm}
\be\l{Boltzlin}
=- i \frac{p_z}{mT}\;f^{}_{\rm GE}\, \exp\left(-i~\omega t\right)  \;
\!\!\int_{-\infty}^\infty \frac{\d k}{2 \pi}\,k\, V_k \;
\exp\left(i k z\right).
\ee

We can present $\delta f(\r,\p,t)$
defined by Eq.~(\ref{dfdef}) in terms of the
sum of two terms
\be\l{phi}
\delta f~\equiv ~ \delta f_{\rm LE}~+~\delta \varphi~.
\ee
The local equilibrium term $\delta f_{\rm LE}$ (see Ref.~\cite{silin})
in Eq.~(\ref{phi}) will
be written as
\bea\l{dfle}
\delta f^{}_{\rm LE}
~=~ f^{}_{\rm GE}\,\left(\frac{\delta n}{n}~+~\frac{p_z \delta u_z}{T}
\right)\;,
\eea
where $\delta n$ and $\delta u_z$
are small deviations of the particle number density
and collective velocity ($|\delta n|/n \ll 1$
and $|\delta u_z|/v^{}_T \ll 1$) from their GE values $n$ and $u_z=0$.
Taking into
account that $ \mbox{St}[f] =0 $
for $f=f_{\rm LE}$, one finds that only  the
term $\delta \varphi$ in Eq.~(\ref{phi})
contributes  to the collision integral.
Thus, one can use the so-called
$\tau$ approximation for the collision integral term
$\delta St[f] $
in Eq.~(\ref{Boltzlin}),
\bea \l{tauapprox}
\delta \mbox{St}[f]~\cong~
-\frac{1}{\tau}~
\delta \varphi ~,
\eea
where $\tau$ is the collision relaxation time.

A particle number and momentum conservation impose the following requirements
\cite{baympeth,heipethrav,review}:
\bea\l{phi-1}
\int d\p\,\delta \varphi~=~0~,~~~~~
\int \d\p\, p_z~\delta \varphi~=~0~.
\eea
For a constant temperature $T$, from these equations one finds
also the energy conservation $\int d\p\,p^2\,\delta \varphi \equiv 0$.
Equations (\ref{Boltzlin})--(\ref{phi-1}) define the required
solution $\delta f$ for the APSWs. In these
important steps of our derivations (\ref{phi})--(\ref{phi-1})
the variations $\delta n$ and
$\delta u_z$, defined through the variations of Eq.~(\ref{p-mom}),
were determined by the moments
of $\delta f^{}_{\rm LE}$ in Eq.~(\ref{dfle}).

The relaxation time $\tau$ can be evaluated {\bf \cite{MGGPprc16}}
within the molecular kinetic
theory as
\be\l{tau}
\tau ~\sim ~ \frac{l}{v^{}_T}~\sim 
\frac{1}{ n \;v^{}_T \sigma}\;,
\ee
%
where $\sigma$ is the cross section for the particle collisions,
$\sigma=\pi d^2$ for the case
of the hard-sphere particles
of a diameter $d$.
%
%
The role of a specific number coefficient in
Eq.~(\ref{tau}) will be discussed later.

\section{Absorbed plane sound-wave solutions
}
\l{sec-denvelres}

The linear BKE (\ref{Boltzlin}) for the distribution function variations
$\delta f(z,\p,t)$ under the oscillating external mean-field
potential $V_{\rm ext}(z,t)$  (\ref{Vext})
will be solved by using the Fourier representation
\bea\l{dfk}
\delta f(z,\p,t)  &=& \int_{-\infty}^{\infty} \frac{\d k}{2\pi}\;
\delta f_k\,\exp\left(-i\omega t + ik z\right)\;,\\
\delta n(z,t)&=&\int_{-\infty}^{\infty} \frac{\d k}{2\pi}\;
\delta n_k\;
\exp\left(-i\omega t + ik z\right)\;, \label{dnk}\\
\delta u_z(z,t)&=&\int_{-\infty}^{\infty} \frac{\d k}{2\pi}\;
\delta u_{z\,k}\;
\exp\left(-i\omega t + ik z\right)\;.\l{duzk}
\eea
Substituting Eqs.~(\ref{dfk})--(\ref{duzk}) into  the
BKE (\ref{Boltzlin}) and
using the definitions
for
$\delta n_k$ and $\delta u_{zk}$
[see Eq.~(\ref{dndudf})], one can
reduce it to the integral
equation for $\delta f_k$ [see Eq.~(\ref{dfk-3})]. This equation
 presents the Fourier plane-wave amplitudes $\delta f_k$, in an algebraic way,
in terms of the Fourier amplitudes $\delta n_k$,
$\delta u_{zk}$, and $V_k$:
\be\l{dfk-1}
\delta f_k=
\frac{i f_{\rm GE}(p)}{\xi - \hat{p}_z}
\left[\frac{c}{\mathcal{K}}\,\left(\frac{1}
{n}\,\delta n_k+
\frac{p_z}{T}\; \delta u_{z\,k}
\right)
-i\,\frac{\hat{p}_z}{T}\;
V_k\,\right]\;,
\ee
where the following notation is introduced:
\be\l{note}
\hat{p}_z \equiv \frac{p_z}{p}\;,~~~~~ c \equiv \frac{\omega}{k\;v^{}_T}\;,~~~~~
\xi\equiv c\left(1+\frac{i}{\mathcal{K}}\right)\;.
\ee
Note that the quantity $c$ is a dimensional (in units of $v^{}_T$)
speed of
the sound wave
with a wave number $k$.

Equations (\ref{phi-1})
connecting $\delta f$  with $\delta n$ and $\delta u_z $
can be rewritten in terms of the Fourier amplitudes as
\be\l{phi-2}
\int \d \p \; \delta f_k(\p) = \delta n_k~,\quad
\frac{1}{m\,n}\int \d \p\; p_z\; \delta f_k(\p)
=\delta u_{zk} ~,
\ee
where $\delta f_k$ are given by Eq.~(\ref{dfk-1}).
Substituting
amplitudes $\delta f_k(\p)$ [Eq.~(\ref{dfk-1})]
into  Eq.~(\ref{phi-2}),
one calculates explicitly the integrals
over $\p$ with the help of
the GE distribution
function $f_{\rm GE}(p)$ given by Eq.\ (\ref{maxwell}).
As shown in Appendix A, this leads to a system of two linear
equations for $\delta n_k$ and $\delta u_{zk}$ in units of $V_k$
[see Eqs.~(\ref{A1}) and (\ref{A2})].
The Fourier components
$\delta f_k$ can then be expressed in terms of the linear
response functions
$\mathcal{D}_k=\delta n_k/V_k$ and
$\mathcal{U}_k=\delta u_{zk}/V_k$ [see
Eq.~(\ref{dndudf})]. Thus, one obtains
\bea\l{dfk-2}
\delta f_k &=&
\frac{i f^{}_{\rm GE}(p)}{\left(\xi - \hat{p}_z\right)\;
\mbox{D}(c,\mathcal{K})}~\nonumber\\
&\times&\left[\frac{c}{\mathcal K}\,
\left(\frac{\alpha_k}{n}+
\frac{p_z}{T}\; \beta_k\,\right)
-\frac{i\hat{p}_z}{T}
\right]\;V_k\;,
\eea
where $\alpha_k$ and $\beta_k$ are defined in
Eq.~(\ref{Cramer})
and $\mbox{D}(c,\mathcal{K})$ is given by Eq.~(\ref{D}).
The APSW distribution function
 $\delta f_k$ is expressed in terms of
these response functions. Thus,
the linearized BKE (\ref{Boltzlin}) is solved
in a simple closed form
for an arbitrary Knudsen parameter $\omega\tau$.

For the particle density $\delta n_k$ and mean velocity $\delta u_{z k}$
components
one obtains the explicit expressions through the
linear response
functions $\mathcal{D}_k$ and $\mathcal{U}_k$,
\be\l{denkuzk}
\delta n_k=\frac{\alpha_k}{\mbox{D}}~V_k~,\qquad
\delta u_{z k}=\frac{\beta_k}{\mbox{D}}~V_k~,
\ee
where $\alpha_k$ and $\beta_k$ and $\mbox{D}$ are given explicitly by
Eqs.~(\ref{Cramer}) and $\mbox{D}$ is given by (\ref{D}).
Taking the integrals (\ref{dnk}) for $\delta n$ and
(\ref{duzk}) for $\delta u_z$ over $k$ with the
Fourier  amplitudes
$\delta n_k$ and $\delta u_{z k}$ [Eq.~(\ref{denkuzk})]
by the residue method, one
notes that
the common poles of both these response functions are
determined by the dispersion equation
\be\l{dispeq}
\mbox{D}\left(c,\mathcal{K}\right)
=0\;.
\ee
These poles $c^{}_0=c_r+i c_i$ correspond
to the collective excitations of the
sound waves.  From Eqs.~(\ref{dispeq}) and (\ref{D})
one finds that
\begin{figure}
\begin{center}
\includegraphics[width=0.49\textwidth]{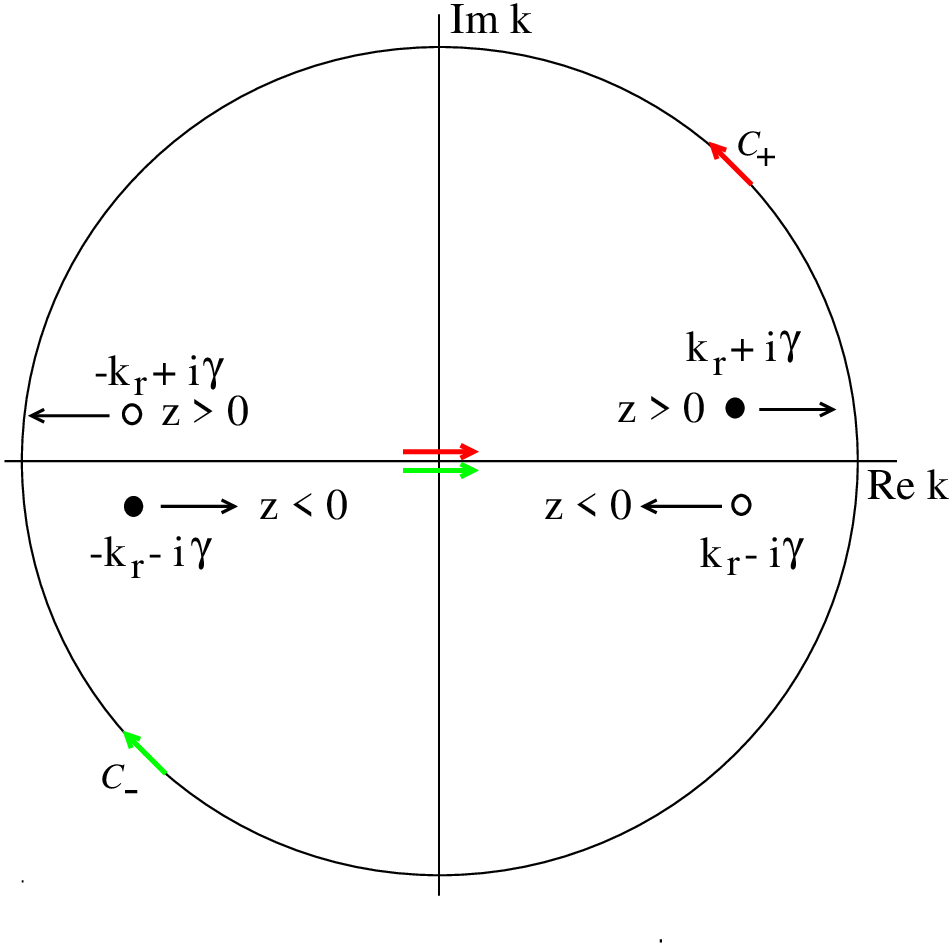}
\end{center}
\vspace{-0.5cm}
\caption{\small{
Schematic representation of the two symmetric poles $k_r+i\gamma$
and $-k_r-i\gamma$ by closed circles
and two symmetric poles $-k_r+i\gamma$
and $k_r -i\gamma$ by open circles
in the complex $k$ plane. Horizontal
arrows near the poles show the wave direction. Also shown are
two closed integration contours
$\mathcal{C}_{+}$ for $z>0$ and $\mathcal{C}_{-}$ for $z<0$ of
the residue method.
}
}
\label{fig1}
\end{figure}
the determinant
$\mbox{D}(c,\mathcal{K})$
is an even function of $c$.
As shown in Fig.~\ref{fig1},
one then obtains, through the relation
(\ref{note}), four
poles in the complex $k$ plane
\be\l{kzp}
k_0=\pm k_{r} \pm i\gamma~,
\ee
 where $k_{r}$ and $\gamma$ are
the absolute values of the sound-wave number
and the absorption coefficient, respectively,
\be\l{krdef}
k_r~=~\frac{\omega}{v_T}~\frac{|c_r|}{c_r^2+c_i^2}\;,~~~~~~
\frac{\gamma}{k_r}~=~\Big|\frac{c_i}{c_r}\Big|~.
\ee
The poles $k_r+i\gamma$ and $-k_r-i\gamma$ are related to the plane waves
moving in the positive direction of the $z$ axis while
$-k_r+i\gamma$ and $k_r-i\gamma$ correspond to the sound waves spreading
in the negative $z$ direction (see Fig.~\ref{fig1}).
Taking, for example, $z>0$,
one can close the
integration
path along the real axis of the complex $k$ plane in
Eqs.~(\ref{dnk}) and (\ref{duzk})
 by adding
the integration contour over
the semicircle of a large radius in
the upper half
of this plane.
The integrand along such a semicircle decreases exponentially
with increasing the radius to infinity.
This integration
in Eqs.~(\ref{dnk}) and (\ref{duzk}) can be performed by
the residue method.
As the result,
one arrives at the
APSW particle density [Eq.~(\ref{dnk})] and velocity field [Eq.~(\ref{duzk})],
which are related to one of the poles $k^{}_0=k_r+i\gamma$
[Eqs.~(\ref{kzp}) and (\ref{krdef})]
for waves moving in the positive direction,
\bea\l{dnusw}
\hspace{-0.5cm}\delta n(z,t)&\!=\!&\!\left(\frac{\alpha_{k}V_{k}}{\d \mbox{D}/\d k}\right)_{k=k^{}_0}
\!\!\!\exp\left[-i\left(\omega t -k_rz\right)-\gamma z\right],~~~\nonumber\\
\delta u_z(z,t)&\!=\!&\!\left(\frac{\beta_{k}V_{k}}{\d \mbox{D}/\d k}\right)_{k=k^{}_0}
\!\!\!\exp\left[-i\left(\omega t -k_rz\right)-\gamma z\right].~~~
\eea
Thus, one obtains the wave number $k_r>0$ and the absorption
coefficient $\gamma>0$ for sound waves spreading in the
positive $z$ axis direction  for $z>0$. Similarly, one finds the contributions
of other poles.

Let us consider the FCR
where $\mathcal{K} \ll 1$.
Taking the asymptotic expansion
of $\mbox{D}(c,\mathcal{K})$ [Eq.~(\ref{D})] in a series over $\mathcal{K}$,
one finds
(see Appendix B)
\bea\l{FCR}
c_r
&\cong& \sqrt{\frac{8}{9\pi}} +a^{}_2 (\omega\tau)^2
+O\left((\omega\tau)^4\right)~,\nonumber\\
\frac{\gamma}{k_r} &\cong& \frac{21\pi-40}{40}\,\omega\tau
+O\left(
(\omega\tau)^3\right)~,
\eea
where $a^{}_2$ is a constant given by
Eq.~(\ref{a0123}).
In the RCR $\mathcal{K} \gg 1$,
one
obtains, from the asymptotic expansion of  Eq.~(\ref{dispeq})
in $1/\omega\tau$
(Appendix C),
\bea\l{RCR}
c_r&\cong& 1~-\frac{1}{(\omega\tau)^2}
+
O\left((\omega\tau)^{-4}\right)~,\nonumber\\
\frac{\gamma}{k_r}&\cong& \frac{1}{\omega\tau}+
O\left((\omega\tau)^{-4}\right)~.
\eea
For small values of
$\mathcal{K}$, or
 $\mathcal{K}^{-1}$, the real part $c_r$
of $c$ is an
even function of $\mathcal{K}$, or
 $\mathcal{K}^{-1}$, i.e., it is
  expanded in even powers, while its imaginary part $c_i$ is expanded
in odd powers.
We emphasize that the FCR (\ref{FCR}) and RCR (\ref{RCR}) limits
were obtained for the same sound velocity $c_r$ and absorption coefficient
$\gamma/k_r$ as obtained, within the LRT, by the numerical solving of the
dispersion equation
(\ref{dispeq}) with the function (\ref{D}).
Notice also that one can obtain terms of the expansion in
powers of $~\omega\tau~$ $~(1/\omega\tau)$ at all
orders by
using the perturbation FCR (RCR) expansion
and applying
the same standard method of indeterminate multipliers
in derivations
of both Appendixes \ref{appB} and \ref{appC}.

\begin{figure*}
\vspace{-0.6cm}
\begin{center}
\includegraphics[width=0.68\textwidth]{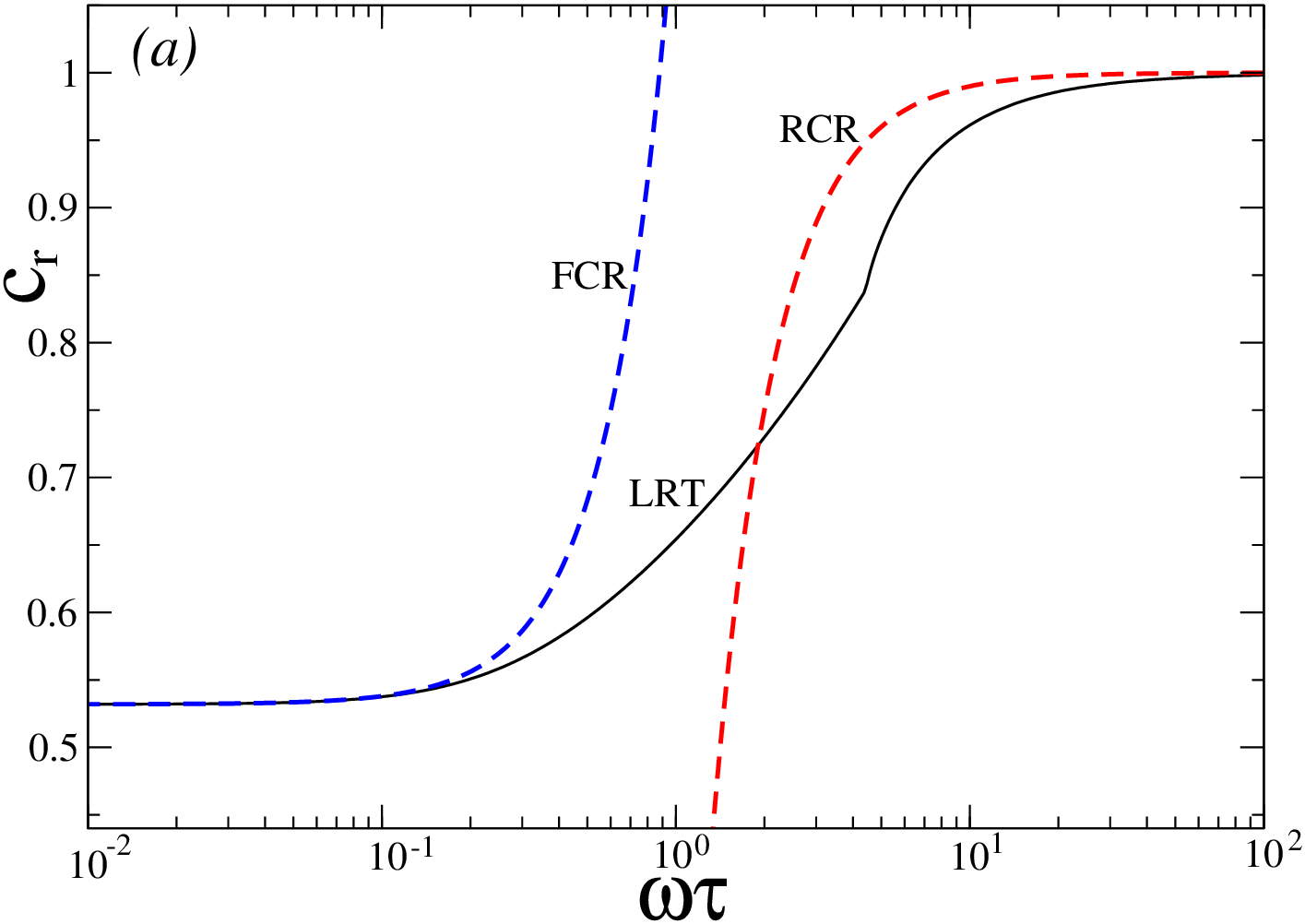}
\end{center}
~

\vspace{0.7cm}
\includegraphics[width=0.68\textwidth]{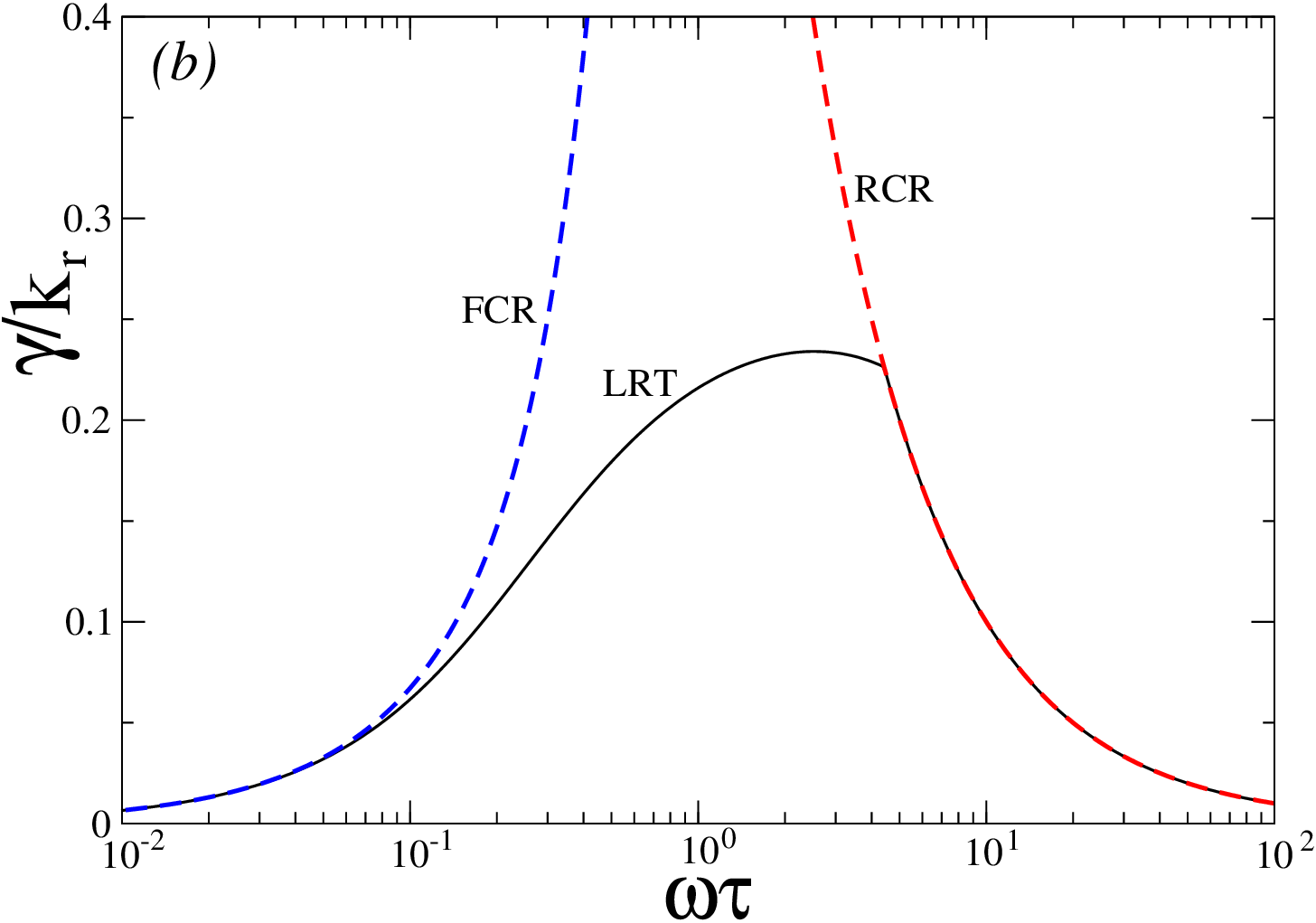}

\vspace{0.2cm}
\caption{\small{
(a) Sound velocity
 $c_r$
in units of the thermal value $v^{}_T$ and (b) scaled
absorption coefficient $\gamma/k_r$ as functions
of the Knudsen parameter $\omega \tau$.
Solid lines show the
nonperturbative LRT solutions.
Dashed lines present
the asymptotic FCR
[Eq.~(\ref{FCR})] and the RCR [Eq.~(\ref{RCR})] approximations.
}
}
\label{fig2}
\end{figure*}

\section{Discussion of the results}
\l{sec-disc}

Figure \ref{fig2}
shows
the sound velocity $c_r$
and scaled absorption coefficient
$\gamma/k_r$
as functions of the Knudsen parameter $\omega\tau$.
The results presented by solid lines
are obtained by
numerically solving the dispersion equation (\ref{dispeq}).
The sound velocity $c_r$ is defined as a dimensionless quantity in units of
$v_T$ given by Eq.~(\ref{vT}).
The quantity $v_T$ is  the thermal particle velocity in
a classical gas defined by
the Maxwell distribution (\ref{maxwell}).
Figure \ref{fig2}(a) demonstrates a nontrivial
$\omega$ dependence of the sound velocity.
 In both limits $\omega\tau \ll 1$ and
$\omega\tau \gg 1$, our numerical results converge to the asymptotic limits
of the FCR and RCR, respectively. These limiting behaviors
corresponding
to  Eqs.~(\ref{FCR}) and
(\ref{RCR}) at leading (quadratic) orders
are shown by the dashed
lines in this figure.
Note that the sound velocity  and scaled absorption coefficients are
presented as universal
functions of the Knudsen parameter $\omega \tau$ and, therefore, the
lines in Fig.2 do not depend on the specific number coefficient in
$\tau$ [Eq.~(\ref{tau})]. However,   their $\omega$ behavior depends 
on this number coefficient such that 
the whole picture
is shifted along the abscissa axis without changing the
shapes of any lines.

The scaled absorption coefficient $\gamma/k_r$ measures how the
amplitude of APSWs decreases after propagating a distance equal to
the wavelength $\lambda$. The APSW amplitude decreases by the factor
$e^{-1}$ with
propagating the distance $\Delta z=1/\gamma $.
In the FCR, for gases at normal conditions, one gets
$\omega\tau\sim 10^{-8}--10^{-5} $ for the audible frequency region.
This gives
 $\Delta z=1/\gamma^{}_{\rm FC} \sim l (\omega \tau)^{-2}\sim 10^5--
 10^{11}$~cm. Thus, the audible
sound wave propagates a distance of several kilometers without essential
absorption in a gas. At the RCR
for $\omega\tau \gg 1$,
the behavior is quite different,
$\Delta z =1/\gamma_{\rm RC}\sim l $, i.e., the propagating length is
of the order
of a mean free path. This quantity remains rather small even
for dilute gases.
Nevertheless, this regime is realized in experiments
\cite{Gr65book,Gr56,Me57}
by using much larger ultrasonic frequencies
and much smaller pressures (particle
densities) of the gas. Note that
these numbers in the FCR case are well known.
We recall these known numbers to stress their great
difference from those in the RCR. Very short propagation lengths of ultrasonic waves
in gases are one of the problems for careful measurements of the absorption
coefficient.

Figure \ref{fig2}(b) shows
the scaled absorption coefficient
$\gamma/k_r$  [Eq.\ (\ref{krdef})].
As can be seen from
this figure, our numerical nonperturbative
results  for the scaled
absorption coefficient shown by
the solid line are in agreement
with both the FCR (\ref{FCR}) and the RCR (\ref{RCR}) asymptotic limit
presented by dashed lines. The
absorption coefficient $\gamma/k_r$  as a
function of the Knudsen parameter
demonstrates  a maximum at $\omega\tau \approx 1$
in the transition from the FCR to the RCR.

The kink in the dependence of the scaled absorption coefficient
$\gamma/k_r$ and of the sound velocity $c_r$ on the Knudsen
parameter is found at $\omega\tau\approx 4.4$,
where the derivatives with respect
to $\omega\tau$ are sharply changed.
This is
obtained in the numerical calculations, which
are carefully
checked within two different numerical schemes. There are two
length scales in the problem: the mean free path of particles
in  a gas $l=v^{}_T\tau$ and the sound
wavelength $\lambda=c_rv_T 2\pi/\omega$.   The kink corresponds to
the $\omega\tau$ point where these two different scales become
approximately equal,
$l\approx \lambda$. The presence of the kink cannot be proved as
a mathematical theorem.
It takes place in the nonperturbative region of $\omega\tau$ values, where no analytical
solutions can be obtained. This resembles a situation similar  to phase transitions
in statistical mechanics. The origin of the kink remains an open problem that deserves
further studies.

\section{Summary}

The kinetic approach for calculations of
the velocity and absorption coefficient for the absorbed
plane sound waves is developed
by solving the linearized Boltzmann kinetic equation with a small external
plane-wave perturbation potential. The solution
is based on the relaxation time approximation to the Boltzmann
collision integral term for  classical dilute gases.
It was explicitly demonstrated
that the mean
particle density and velocity responses of the system,  $\delta n$ and
$\delta u_{z}$, to a small
external potential $V_{\rm ext}$
determine
the distribution function
response $\delta f$ within
the linear response theory. We obtained also explicitly
 the sound excitations
of the particle density $\delta n(z,t)$ and velocity field
$\delta u_{z}(z,t)$ in terms of the collective poles of the
dispersion equation and found their
structure.

The nonperturbative numerical solution to this equation
is found for the sound velocity
and absorption coefficient as functions of the Knudsen parameter
$\omega\tau$ beyond  both  the  frequent-collision
and rare-collision regime
approximations. This numerical solution agrees with the asymptotic expansions in both 
FCR and RCR.
We found a dramatic change of the scaled
absorption coefficient  in the transition region
between the
frequent-collision
and rare-collision regimes: a maximum of $\gamma/k_r$
at $\omega\tau \approx 1$.

The  RCR in the kinetic theory is a subject that
is not restricted
only to the ultrasonic waves
in classical gases.
A strong difference between the FCR and RCR for sound waves means that
the shear viscosity
behaves very differently in these two regimes (see the discussion of this
point in Ref.~[35]).
Other kinetic coefficients, i.e., the thermal conductivity and
the diffusion coefficient,
should also behave
in a different way for the FCR and RCR.

One possible application of the kinetic theory in the RCR is high-energy
nucleus-nucleus collisions.
The intermediate stage of these collisions is often described by the
hydrodynamical approach. The hydrodynamic
description should be stopped at some stage (the so-called freeze-out procedure).
After such a stage
the system is usually considered as  that of
freestreaming particles. At this post-freeze-out stage,
however, the particle collisions still occur and final momentum spectra are influenced by these
collisions. This stage is the RCR of the kinetic models. The mean free path
of particles flying away becomes larger than the system
size.
\begin{acknowledgments}
We thank D.V.\ Anchishkin, V.P.\ Gusynin,
V.M.\ Kolomietz, B.I.\ Lev, V.A.\ Plujko,
S.N.\ Reznik, A.I.\ Sanzhur, Yu.M.\ Sinyukov, and A.G.\ Zagorodny for
many fruitful discussions.
The work of A.G.M. was
supported by the Fundamental Research Program
``Nuclear matter in extreme conditions''
at the Department of Nuclear Physics and Energy of the National
Academy of Sciences of Ukraine through Grant No. CO-2-14/2017.
The work of M.I.G. was supported
by the Program of Fundamental Research of the Department of
Physics and Astronomy of the National Academy of Sciences of Ukraine.

\end{acknowledgments}

\appendix

\section{Method of the linear response}\l{appA}

Using the linearization procedure in Eq.~(\ref{p-mom}) and Fourier
transformations (\ref{dfk})--(\ref{duzk}), one has
\bea\l{dndudf}
\delta n_k &=&\int \d\p\;\delta f_k(\p) \equiv
\mathcal{D}_k\;V_k\;,\nonumber\\
\delta u_{z k} &=&\frac{1}{mn}\int \d\p\;p_z\delta f_k(\p)
\equiv
\mathcal{U}_k\;V_k\;,
\eea
where $\mathcal{D}_k$
and $\mathcal{U}_k$
are the corresponding response
functions \cite{kubo,zubarev,hofmann,baympeth,heipethrav,magkohofsh,review}.
Substituting
Eq.~(\ref{dndudf}) into Eq.~(\ref{dfk-1}), one can represent  it in the explicit
form of the integral
equation for $\delta f_k(\p)$:
\bea\l{dfk-3}
\delta f_k(\p)&=&
\frac{i f_{\rm GE}(p)}{\xi - \hat{p}_z}\;
\left[\frac{c}{\mathcal{K}}\,\left(\frac{1}
{n}\,\int\d\p^\prime~\delta f_k(\p^\prime)\right.\right.\nonumber\\
&+&\left.\left.
\frac{p_z}{mnT}\; \int\d\p^\prime\;p_z^\prime~\delta f_{k}(\p^\prime)
\right) -i\,\frac{\hat{p}_z}{T}\;
V_k\,\right]\;.
\eea
This equation can be solved in terms of
the response functions $\mathcal{D}_k$
and $\mathcal{U}_k$ [Eq.~(\ref{dndudf})]. From Eq.~(\ref{phi-2}) with the
help of Eq.~(\ref{dfk-1}),
one gets the two
linear equations
\bea\label{A1}
A_1\,\mathcal{D}_k
+B_1\,\mathcal{U}_k
=C_1
~,
\\
A_2\,\mathcal{D}_k
+B_2\,\mathcal{U}_k
=C_2\,
~,
\label{A2}
\eea
where
\bea\l{abc}
 A_1&=&
\left[i\,c\left(Q_1+1\right)/\left(\mathcal{K}\xi\right)-1\right]/n~,
 \nonumber\\
B_1&=&
4i\,c\,Q_1/\left(\sqrt{\pi}~\mathcal{K}~v^{}_T\right)~,\nonumber\\
C_1&=&-Q_1/T~,\nonumber\\
 A_2&=&
2i\,c~Q_1/\left(\sqrt{\pi}~\mathcal{K}~n\right)~,\nonumber\\
B_2&=&
\left(3i\xi\,cQ_1/\mathcal{K}-1\right)/v^{}_T~,\nonumber\\
C_2&=&-2\xi Q_1/\left(\sqrt{\pi}~T\right)~,
\eea
and $Q_1$ is the Legendre function of second kind,
\be\l{Q1}
Q_1(\xi)~=~\frac{\xi}{2}\,\ln\left(\frac{\xi+1}{\xi-1}\right)~-~1~.
\ee

Solving Eqs.~(\ref{A1}) and (\ref{A2}), one obtains
\bea\l{Cramer}
\mathcal{D}_k
&=&
\frac{B_2C_1 -B_1C_2}{\mbox{D}}
\equiv\frac{\alpha_k}{\mbox{D}}\,
~,\nonumber\\
\mathcal{U}_k&=&
\frac{A_1C_2 -A_2C_1}{\mbox{D}}
\equiv\frac{\beta_k}{\mbox{D}}\,
~,
\eea
where $\mbox{D}$ is the main determinant of the linear system of
Eqs.~(\ref{A1}) and
(\ref{A2}),
\bea\l{D}
\mbox{D}&=&\mbox{D}(c,\mathcal{K})
=A_1B_2-A_2B_1\nonumber\\
&=&\frac{1}{nv^{}_T}\left\{\left[
\frac{i\,c}{\xi\,\mathcal{K}}\;\left(1+Q_1\right)
-1\right]
\left(3 i\,\frac{c}{\mathcal{K}}\, \xi \,
Q_1-1\right) \right.\nonumber\\
&+&\left.\frac{8}{\pi}\;\left(\frac{c\, Q_1}{\mathcal{K}}\right)^2\right\}.
\eea
Substituting Eq.~(\ref{Cramer}) into Eq.~(\ref{dfk-1}) with definitions
(\ref{dndudf}), one arrives at the final solution (\ref{dfk-2}) in terms
of the response functions $\mathcal{D}_k$
and
$\mathcal{U}_k$
with their explicit expressions (\ref{Cramer}).

\section{FCR derivations}\l{appB}

By using the universal method of
indeterminate coefficients \cite{brooksyk}
in the case of the FCR
perturbation expansion over small $\mathcal{K}$,
\be\l{cpertFCR}
c=a^{}_0 + a^{}_1~\mathcal{K} + a^{}_2~\mathcal{K}^2 +
a^{}_3~\mathcal{K}^3 + a^{}_4~\mathcal{K}^4 + \cdot\cdot\cdot ~,
\ee
one can solve the dispersion equation (\ref{dispeq}) with
Eq.~(\ref{D}) for the determinant $\mbox{D}$.
Substituting this perturbation series
into the expansion of the function $\mbox{D}(c,\mathcal{K})$
[Eq.~(\ref{D})] over $\mathcal{K}$ at a given $c$
(multiplied by a nonzero factor, e.g., $c^6/\mathcal{K}^2$
at the $\mathcal{K}^4$ order), one obtains the series that is an identity
in powers of $\mathcal{K}$.
Equaling the coefficients of this series at each power of
$\mathcal{K}$,
one arrives at the system of equations for the coefficients
$a_n$ ($n=0,1,2,\ldots $):
\be\l{a0FCR}
a^{2}_0-\frac{8}{9\pi}=0\;,
\ee
\be\l{a1FCR}
\left(135\pi~ a_0^2-80\right) a^{}_1 +  45~i\pi~ a_0^3+
i(21\pi-80) a^{}_0=0\;,
\ee
\bea\l{a2FCR}
&10\left(27\pi~a^2_0 - 16\right) a^{}_0 a^{}_2+
4i\left(135\pi~ a^2_0+42\pi-160\right) a^{}_0 a^{}_1 \nonumber\\
&+
\left(675 \pi~a^2_0-240\right) a^2_1
-\left(135\pi~a^2_0 +168\pi-400\right) a^2_0\nonumber\\
&+
48-9\pi=0\;,
\eea
and so on. We omitted common coefficients proportional to positive
integer powers of $a_0$
to get the nonzero solutions. Solving consequently these equations step by step
(for nonzero solutions),
one obtains all coefficients.

For instance, at fourth-order terms, one obtains the polynomial equation
of the sixth order with respect to $c$. Therefore,
one finds three pairs of
analytical solutions
$c_n$, $n=1,2,\ldots ,6$, namely $c^{}_2=-c^{}_1$,
$c^{}_4=-c^{}_3$, and $c^{}_6=-c^{}_{5}$. The solutions of three roots,
$a^{}_1$, $a^{}_3$ and $a^{}_5$ can be presented by the Cardan formulas.
For definiteness,
taking the roots related to the positive solution for $a_0$ of
Eq.~(\ref{a0FCR}), from Eqs.~(\ref{a0FCR})--(\ref{a2FCR}) one obtains
\bea\l{a0123}
a^{}_0&=&\sqrt{\frac{8}{9\pi}}
\;,\qquad a^{}_1=
i\frac{40 - 21 \pi}{30 \sqrt{2 \pi}}
\;,~~\nonumber\\
a^{}_2&=& \frac{423 \pi^2 - 240 \pi -1600}{1200 \sqrt{2 \pi}}
\;,~~
\nonumber\\
a^{}_3&=&i\frac{30429 \pi^3- 44910 \pi^2- 16800 \pi- 112000}{84000 \sqrt{2 \pi}},~~
\eea
and so on.
Two other roots $c^{}_3$ and $c^{}_{5}$ are expansions
over powers of
$\mathcal{K}$ starting from the first-order term, which is proportional
to $\mathcal{K}$.
They are complex conjugated, $c^{}_5=c^{\ast}_3$.
For our purpose of the APSW description
we have to select the root $c^{}_1$ of the
dispersion equation
with a finite
constant in the limit $\mathcal{K}\rightarrow 0$ as the physical
solution.

The real part $c_r$ of the complex roots
of Eq.~(\ref{cpertFCR}) is expanded in even powers of $\mathcal{K}$ while
its imaginary part $c_{i}$ is an expansion in  odd powers of $\mathcal{K}$.
Splitting Eq.~(\ref{cpertFCR}) with found coefficients $a^{}_n$
[Eq.~(\ref{a0123})] into the real $c_r$ and imaginary $c_i$ parts of
$c$, one finds
\bea\l{FCRrootr}
c_r(\mathcal{K})&=& a^{}_0 + a^{}_2\mathcal{K}^2 +
O\left(\mathcal{K}^4\right)\;,
\nonumber\\
c_i(\mathcal{K})&=& a^{}_1 \mathcal{K} +
a^{}_3 \mathcal{K}^3 + O\left(\mathcal{K}^5\right)\;.
\eea
Therefore, according to Eq.~(\ref{krdef}),
for the scaled absorption coefficient
$\gamma/k_r$, one
obtains the following FCR result up to fifth order terms:
\be\l{FCRgk3}
\frac{\gamma}{k_r}=\frac{21\pi-40}{40}\;\mathcal{K}
+q^{}_3\mathcal{K}^3
+ O(\mathcal{K}^5)\;,
\ee
where $q^{}_3$ is
 an analytically given (rather bulky) number $q^{}_3=-3.73421\ldots~$~.
Using Eqs.~(\ref{FCRrootr}) and (\ref{FCRgk3}), one finally arrives at
the main terms of Eq.~(\ref{FCR}).

\section{RCR derivations}\l{appC}

In the RCR case, let us start with the perturbation expansion of
solutions of the dispersion equation  (\ref{dispeq}) with Eq.~(\ref{D})
for the determinant function $\mbox{D}(c,\mathcal{K})$ over
a small Knudsen parameter $1/\mathcal{K}$,
\be\l{croot4}
c=a_0^{} + a_1^{}/\mathcal{K} + a_2^{}/\mathcal{K}^2 +
a_3^{}/\mathcal{K}^3 + a_4^{}/\mathcal{K}^4 + \cdot\cdot\cdot~,
\ee
where $a_n$ are new indeterminate coefficients.
For simplicity of the presentation,
up to third-order terms of the expansion
of $\mbox{D}(c,\mathcal{K})$ over $1/\mathcal{K}$,
one obtains
\bea\l{Dfunexp4}
&nv^{}_T\mbox{D}(c,\mathcal{K})=1 - \frac{ic}{2\mathcal{K}}\;
\left[-6 c +(1+3c^2) \ln\left(\frac{c+1}{c-1}\right)\right]\nonumber\\
&-\frac{c^2}{4(c^2-1)\pi \mathcal{K}^2}\left\{8 \left[4-\pi+c^2(3\pi-4)\right]
\right.\nonumber\\
&-\left.2c(c^2-1)(9\pi-16)~\ln\left(\frac{c+1}{c-1}\right)\right.\nonumber\\
&+\left.
c^2(c^2-1)(3\pi-8)~\ln^2\left(\frac{c+1}{c-1}\right)\right\} +
\mbox{O}\left(\frac{1}{\mathcal{K}^3}\right)\;.
\eea
Substituting the expansion (\ref{croot4})
with unknown coefficients $a_n$ ($n=0,1,\ldots ,$)
into the dispersion equation (\ref{dispeq}) with Eq.~(\ref{Dfunexp4})
for $\mbox{D}$
and setting zero the expressions
at any given order in $1/\mathcal{K}$ of the obtained identity,
 one finds the nonlinear
system of equations with respect to these constants
$a_n$.
 Starting from the
linear approximation in $1/\mathcal{K}$, one obtains
\bea\l{eqa0}
&\mbox{D}_1(a^{}_0,\mathcal{K})\equiv 1 - ia^{}_0/(2\mathcal{K})\;
\nonumber\\
&\times
\left[(1+3a^2_0) \ln\left(\frac{a^{}_0+1}{a^{}_0-1}\right)
- 6 a^{}_0\right]=0\;.
\eea
At the second order, one
finds the equation for $a^{}_1$ as function of
$a^{}_0$,
\bea\l{eqa0a1}
&\mbox{D}_2(a^{}_0,a^{}_1)\equiv
4a^{}_0\left[2a^{}_0(4-\pi)+a^{3}_0(6\pi-8)\right.\nonumber\\
&+i \left.\pi~ a^{}_1(5 -9 a^{2}_0)\right]
-2(a^{2}_0-1)\left[a^{3}_{0}(9 \pi-16)\right.\nonumber\\
 &-\left.i \pi \left(1+9a^{2}_0\right)a^{}_1\right]
\ln\left(\frac{a^{}_0+1}{a^{}_0-1}\right)\nonumber\\
 &+a^{4}_0(a^{2}_0-1)(3\pi-8)
\ln^2\left(\frac{a^{}_0+1}{a^{}_0-1}\right)=0\;,
\eea
and so on. The next equation having the structure
$\mbox{D}_3(a^{}_{0},a^{}_{1},a^{}_{2})=0~$ gives $a^{}_2$
as function of
$a^{}_0$ and $a^{}_1$.
 Eq.~(\ref{eqa0}) is a
transcendent equation for only one variable $a^{}_0$.
Any given $n$th equation of this system $D_n(a^{}_0, a^{}_1,...,a^{}_{n-1})=0$
is linear with respect to the last argument $a^{}_{n-1}$.
The last coefficient $a_{n-1}$  can easy be
found analytically,
as a function of all other (with smaller subscripts)
variables.
This determines the iteration perturbation procedure
to obtain all of the coefficients $a_n$ in Eq.~(\ref{croot4}).

To get explicitly
expansions for $a_n$ over $\mathcal{K}$, one notes
that the solution of the first equation (\ref{eqa0}) with respect to
$a^{}_0$ converges asymptotically to one  at $\mathcal{K} \to \infty$.
Substituting this solution into expressions for
$a_n$ ($n\geq 1$) in the
remaining equations, one finally obtains the asymptotic series
\be\l{RCcas}
c=\frac{\mathcal{K}}{i+\mathcal{K}}=1 - \frac{i}{\mathcal{K}} -
 \frac{1}{\mathcal{K}^{2}}+\frac{i}{\mathcal{K}^{3}} +
O\left(\frac{1}{\mathcal{K}^{4}}\right)\;.
\ee
The first two (linear) terms were obtained in Ref.~\cite{MGGpre17}
and are used here for the numerical calculations.
Separating the real and imaginary parts from Eq.~(\ref{RCcas}), one gets
\bea\l{RCcrgkas}
c_r&=&1-\frac{1}{\mathcal{K}^{2}} + O\left(\frac{1}{\mathcal{K}^{4}}\right)\;,
\nonumber\\
 c_i&=&-\frac{1}{\mathcal{K}} + \frac{1}{\mathcal{K}^3}+
O\left(\frac{1}{\mathcal{K}^{5}}\right)\;.
\eea
Thus, from this equation one finally
arrives at Eq.~(\ref{RCR}).

\end{document}